# Déploiement des Modèles ARX pour la Prévision Thermique dans les Cartes d'Électronique de Puissance à Semi-Conducteurs WBG

Mohammed Riadh BERRAMDANE[1-2*], Alexandre BATTISTON[1], Michele BARDI[1], Nicolas BLET[2], Benjamin REMY[2], Matthieu URBAIN[2]

[1] IFP Energies nouvelles, 1 et 4 avenue de Bois-Préau, 92852 Rueil Malmaison, France
Institut Carnot IFPEN Transports Energie

[2] Laboratoire Energies et Mécanique Théorique et Appliquée – Université de Lorraine, Centre National de la Recherche Scientifique
2 Av. de la Forêt de Haye, 54500 Vandœuvre-lès-Nancy

**RESUME** – Face aux défis de la gestion thermique des semi-conducteurs WBG, cette étude valorise l'utilisation de modèles paramétriques ARX, qui offrent des prédictions précises de température sans nécessiter une compréhension approfondie des disparités d'épaisseur ou des propriétés physiques des matériaux, se basant uniquement sur des mesures expérimentales. Ces modèles paramétriques se présentent comme une alternative fiable aux simulations FEM et aux modèles thermiques conventionnels, simplifiant significativement l'identification du système tout en garantissant une grande précision des résultats.

**ABSTRACT** – Facing the thermal management challenges of Wide Bandgap (WBG) semiconductors, this study highlights the use of ARX parametric models, which provide accurate temperature predictions without requiring detailed understanding of component thickness disparities or material physical properties, relying solely on experimental measurements. These parametric models emerge as a reliable alternative to FEM simulations and conventional thermal models, significantly simplifying system identification while ensuring high result accuracy.

**MOTS-CLES** – Modèles paramétriques ARX, Semi-conducteurs à large bande interdite (WBG), Identification de système, Modélisation thermique, Electronique de puissance.

## 1. Introduction

L'introduction des semi-conducteurs WBG dans l'électronique de puissance marque une évolution significative, posant notamment des défis de gestion thermique. En raison de leur fonctionnement à haute fréquence et sous fort courant, ces composants sont sujets à des densités de pertes importantes, provoquant une hausse critique de la température de jonction [1][2]. Leur faible épaisseur (de quelques um), ajoutée à celle des substrats électroniques sur lesquels ils sont appliqués, rend les simulations thermiques particulièrement ardues. Cette situation est compliquée par la difficulté d'accéder à des données précises sur les propriétés physiques des matériaux, ajoutant un niveau de complexité à leur étude.

Pour répondre à ces défis, sans recourir à des analyses complexes liées aux propriétés physiques des matériaux ou à l'épaisseur des composants, cette étude se concentre sur l'application des modèles autorégressifs à variables exogènes (ARX). Ces modèles offrent une méthode précise et directe pour estimer la température, basée exclusivement sur des données expérimentales, simplifiant significativement l'approche de modélisation thermique pour les semi-conducteurs WBG.

## 2. L'Approche ARX (AutoRegressive with eXogenous inputs model)

### 2.1 Présentation des ARX

Ljung [4] décrit que, dans le cadre théorique le plus général des modèles polynomiaux, cinq polynômes A, B, C, D et F sont identifiés. Dans le cas simplificateur d'un modèle SISO (single input single output), on a :

$$A(q^{-1})y[k] = \frac{B(q^{-1})}{F(q^{-1})} q^{-n_r} u[k] + \frac{C(q^{-1})}{D(q^{-1})} e[k] \qquad (1)$$

Dans ce modèle, $y$, $u$ et $e$ représentent respectivement la sortie du système, son entrée, et une perturbation extérieure souvent supposée être un bruit blanc pour simplifier. Pour modifier le signal de perturbation, on peut ajuster les polynômes

C et D. Habituellement, il existe un retard, ou effet de latence, entre l'entrée et la sortie du système, que l'on peut représenter par un décalage $n_r$ dans le modèle pour refléter cette dynamique temporelle.

Avec cinq polynômes disponibles, on pourrait théoriquement concevoir 32 structures différentes en choisissant d'inclure ou non chaque polynôme (considérant qu'un polynôme non utilisé est égal à 1). Cependant, les recherches dans ce domaine ont révélé un nombre restreint de structures classiques, moins d'une dizaine, qui sont recensées et organisées en deux sous-catégories principales : celles basées sur l'erreur d'équation et celles basées sur l'erreur de sortie (Tableau 1).

**Tableau 1. Structures classiques de modèles paramétriques polynomiaux.**

| Type de modèle | Polynômes | Nom de la structure |
|---|---|---|
| Erreur d'équation | B | Réponse impulsionnelle |
| | AB | ARX |
| | ABC | ARMAX |
| | AC | ARMA |
| | ABD | ARARX |
| | ABCD | ARARMAX |
| Erreur de sortie | BF | OE |
| | BFCD | BJ (Box-Jenkins) |

Pour justifier le choix des modèles ARX par rapport aux autres structures du tableau, on peut s'appuyer sur leurs propriétés uniques de correspondance avec la réponse impulsionnelle, qui reflète étroitement la physique du système étudié. Selon Loussouarn et al., 2018 [3], les modèles ARX sont particulièrement pertinents car ils peuvent être convertis en modèles convolutifs purs, capturant ainsi la réponse impulsionnelle du système. Cette capacité à modéliser directement la dynamique physique sous-jacente offre une représentation précise et permet une interprétation claire des phénomènes, ce qui peut être crucial pour la compréhension et l'optimisation des performances des systèmes d'électronique de puissance à base de composants WBG.

Dans le cas des ARX, u et e ont en commun le polynôme A au dénominateur (figure 1), ainsi la perturbation fait partie de la dynamique identifiée. Ainsi, la forme discrétisée développée de l'équation donne :

$$y[k] = -\sum_{i=1}^{n_a} a_i y[k-i] + \sum_{i=1}^{n_b} b_i u[k-i-n_k+1] + e[k] \qquad (2)$$

La variable $y[k]$ représente la valeur de la variable y à l'instant $k * \Delta t$, où $\Delta t$ est un pas de temps constant. Les termes $n_a$, $n_b$, et $n_k$ sont des entiers qui définissent respectivement l'ordre autorégressif, l'ordre exogène, et le délai d'entrée-sortie dans un modèle de système dynamique. Les paramètres $a_i$ et $b_i$ sont les coefficients autorégressifs et exogènes du modèle à identifier.

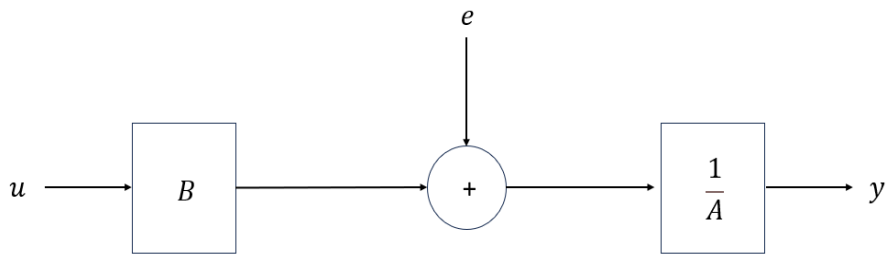

**Figure 1 : Structure ARX**

## 2.2 Utilisation de la modélisation ARX pour estimer les températures des systèmes d'électronique de puissance

Dans le contexte de l'électronique de puissance, et plus précisément pour la modélisation thermique d'une puce électronique, nous pouvons conceptualiser le problème en termes de modélisation ARX, où l'excitation $u$ est représentée par exemple par la puissance P dissipée par la puce et la sortie $y$ est la température T que nous cherchons à prédire.

$$T[k] = -\sum_{i=1}^{n_a} a_i T[k-i] + \sum_{i=1}^{n_b} b_i P[k-i-n_k+1] + e[k] \qquad (3)$$

Pour déterminer les paramètres inconnus $a_i$ et $b_i$ du modèle ARX, nous pouvons reformuler l'équation du système sous une forme matricielle (Eq. (4)) qui lie les vecteurs de température et de puissance à travers une série de coefficients qui doivent être estimés.

$$\begin{bmatrix} T[1] \\ T[2] \\ \vdots \\ T[N] \end{bmatrix} = \begin{bmatrix} T[0] & \cdots & T[-n_a+1] & P[0-n_k] & \cdots & P[-n_k-n_b+1] \\ T[1] & \cdots & T[-n_a+2] & P[1-n_k] & \cdots & P[1-n_k-n_b+1] \\ \vdots & \ddots & \vdots & \vdots & \ddots & \vdots \\ T[N-1] & \cdots & T[N-n_a] & P[N-1-n_b] & \cdots & P[N-n_k-n_b] \end{bmatrix} \cdot \begin{bmatrix} -a_1 \\ \vdots \\ -a_{n_a} \\ b_1 \\ \vdots \\ -b_{n_b} \end{bmatrix} + \begin{bmatrix} e[1] \\ e[2] \\ \vdots \\ e[N] \end{bmatrix} \qquad (4)$$

Dans notre étude, les données intégrant les variations de température et de puissance sont synthétisées dans une matrice simplifiée $\Phi$, et le vecteur $\Omega$ représente les paramètres inconnus à déterminer (Eq. (5)). Pour identifier les valeurs optimales de $\Omega$, la méthode des moindres carrés est employée. Cette approche est reconnue pour sa capacité à réduire au minimum l'écart entre les mesures réelles et les prédictions du modèle (Eq. (6)).

$$T = \Phi.\Omega + e \qquad (5)$$

$$\widehat{\Omega} = \left(\Phi(T,P)^T \Phi(T,P)\right)^{-1} \Phi(T,P)^T T \qquad (6)$$

## 2.3 Implémentation du Modèle ARX pour la Prédiction Thermique des cartes électroniques

### 2.3.1 Principe

Pour calibrer notre modèle ARX, nous soumettons le semi-conducteur à un profil de puissance menant à sa température maximale puis à son retour à l'équilibre thermique. Cette séquence sert d'entrée pour notre modèle $P_e[k]$, avec la réponse en température comme sortie $T_e[k]$. Une étape de validation suit, utilisant un nouveau profil de puissance $P_v[k]$, différent du premier, qui sollicite la puce sur plusieurs points de fonctionnement. Les prédictions de température de notre modèle $T_v[k]$ sont alors comparées aux températures attendues, permettant d'évaluer sa capacité à généraliser et à prédire précisément les températures. Il est à noter que d'autres variables, telle que la température de la semelle, peuvent également être intégrées si besoin en tant qu'entrées supplémentaires, offrant ainsi une modélisation plus précise.

Le processus de développement de notre modèle ARX commence par le traitement des données d'entraînement, où nous ajustons les données de température et de puissance en soustrayant la température ambiante pour nous concentrer sur la variation de température de la puce.

Nous définissons ensuite les intervalles des paramètres autorégressifs $n_a$ et exogènes $n_b$, allant de 1 jusqu'aux valeurs maximales prédéfinies $N_a$ et $N_b$, et le délai $n_k$ variant de 0 à $N_k$. Pour chaque triplet possible de ces paramètres, nous calculons une solution $\Omega$ et sélectionnons le triplet qui offre le meilleur ajustement (fit,Eq.(7)) entre la température prédite $T_{ARX\_e}$ et la température réelle observée $T_e[k]$.

Après avoir choisi le modèle initial, nous entamons la phase de validation. Durant cette étape, nous appliquons le modèle sélectionné pour prédire la température $T_{ARX\_v}$ en fonction d'une nouvelle série de données d'entrée. Cela nous permet de vérifier si le modèle est capable de généraliser au-delà des données d'entraînement et d'éviter le surajustement en se basant uniquement sur le fit, qui pourrait amener à un bon ajustement sur les données d'entraînement mais une performance médiocre sur les données de validation. Si nous identifions un surajustement ou un mauvais ajustement durant la validation, nous revenons à l'étape de définition des paramètres, ajustant le nombre de ces derniers pour améliorer la généralisation du modèle (figure 2). Cette boucle itérative est guidée à la fois par la qualité de l'ajustement ou par le critère d'information d'Akaike (AIC), nous aidant à équilibrer la complexité du modèle et la précision de la prédiction.

$$\widehat{f}_t = 100 \left(1 - \frac{|y - \hat{y}|}{|y - \text{mean}(y)|}\right) \qquad (7)$$

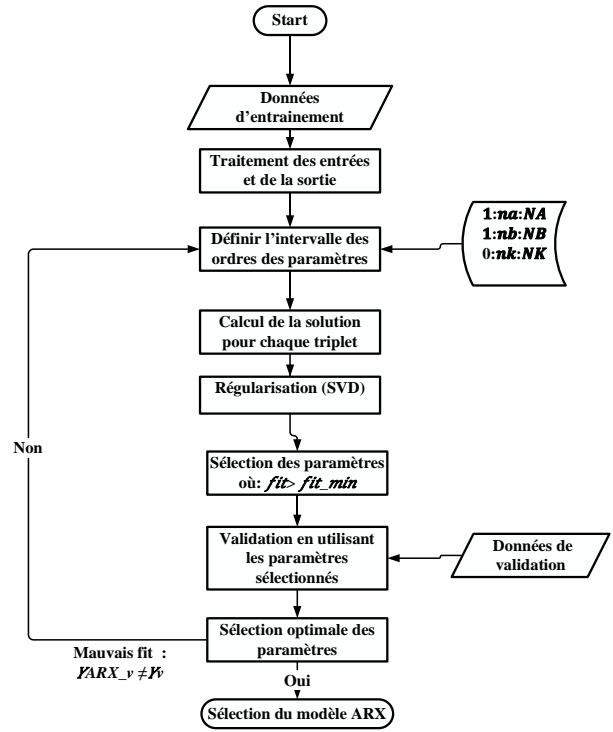

**Figure 2 : Processus Itératif de Développement et de Validation du Modèle ARX**

### 2.3.2   Procédure de régularisation (SVD)

Dans certains cas, le problème d'inversion de la matrice $\Phi$ peut devenir difficile à résoudre avec précision en raison de l'aggravation du conditionnement de $\Phi$. Lorsque le bruit est présent et important, le problème devient plus instable. Concrètement, cela signifie que la matrice $\Phi$ devient difficile à inverser à cause des coefficients de faibles valeurs, ce qui a pour effet d'amplifier le bruit dans les données de température T. Cette situation peut violer les critères de stabilité et fausser les résultats. Une régularisation est donc employée dans ce cas là pour résoudre le problème d'inversion et assurer que les solutions demeurent fiables malgré le bruit.

Dans notre cas, nous appliquons la décomposition en valeurs singulières (SVD) pour la régularisation de la matrice $\Phi$ [5]. Après la décomposition, $\Phi$ est exprimée comme le produit de trois matrices : U, $\Sigma$ et $V^T$, où U et V sont des matrices orthogonales contenant les vecteurs propres, et $\Sigma$ est une matrice diagonale composée des valeurs singulières de $\Phi$. Ces valeurs singulières représentent l'importance de chaque vecteur propre dans la constitution de la matrice $\Phi$.

$$\Omega = U\, \Sigma\, V^T \qquad (8)$$

$$\Sigma = \begin{bmatrix} \sigma_1 & 0 & 0 & 0 & 0 \\ 0 & \sigma_2 & 0 & 0 & 0 \\ 0 & 0 & \ddots & 0 & 0 \\ 0 & 0 & 0 & \sigma_r & 0 \end{bmatrix} \qquad (9)$$

Où les r valeurs singulières de X ont la relation suivante :   $\sigma_1 \geq \sigma_2 \geq \cdots \geq \sigma_r > 0$.

Les valeurs singulières dans la matrice $\Phi$ sont classées par ordre décroissant, celles qui ont le point le plus fort représentent les informations essentielles du signal. À l'inverse, les valeurs singulières plus faibles tendent à correspondre au bruit. En ajustant ces dernières à zéro, on peut filtrer le bruit et affiner la matrice $\Phi$ en vue d'un traitement plus clair du signal. Cette technique est une étape cruciale pour améliorer la précision de nos analyses, comme détaillé dans l'algorithme suivant que nous avons utilisé :

**Entrée :** Matrice $\Phi$.

**Sortie :** Matrice $\Phi_{new}$.

**Procédure :**

1. **Obtenir** U, V, et $\Sigma$ via la décomposition en valeurs singulières (SVD).

2. **Définir** un ensemble de seuils $\Theta$ de n valeurs et initialiser le meilleur ajustement $\text{fit}_{best}$.

3. **Pour chaque** seuil $\Theta_n$ dans $\Theta$ :

   - **Modifier** $\Sigma$ en fixant à zéro toutes les valeurs singulières inférieures à $\Theta_n$ (si $\sigma_i < \Theta_n$ alors $\sigma_i = 0$).
   - **Reconstruire** la matrice $\Phi_n$ à partir des valeurs singulières modifiées.
   - **Évaluer** l'ajustement du vecteur $\Phi_n$ et mettre à jour $\text{fit}_{best}$ et $\Phi_{new}$ si nécessaire (figure 2 : bloc 5 → bloc 4).

4. **Retourner** le vecteur $\Phi_{new}$ correspondant au meilleur ajustement trouvé (figure 2 : bloc 5 → bloc 6).

## 3. Résultats Expérimentaux

Dans notre configuration expérimentale, nous utilisons deux cartes électroniques, chacune contenant quatre puces SiC UJ4SC075011B7S disposées en deux paires parallèles, formant ainsi un bras d'onduleur. Ces cartes sont connectées pour créer un pont en H, où la distance est variable entre les puces sur les deux cartes ; on s'attend à observer des variations de température en fonction de cette distance (figures 3/4).

L'élément clé de notre expérimentation est l'utilisation de capteurs de température à coefficient de température négatif (CTN) placés sur chaque carte à coté de la puce. Ces capteurs fournissent des données de température, essentielles pour l'utilisation du modèle ARX dans l'identification des caractéristiques thermiques des puces.

Pour simuler des conditions opérationnelles réalistes, nos composants SiC sont soumis à un courant maximal de 60A et commandés en PWM, avec une charge inductive dans une configuration de pont en H permettant l'inversion de la polarité (figure8). Cette configuration utilise une source de tension continue de 20V et une fréquence de commutation de 15 kHz.

En plus de la configuration expérimentale décrite, nous appliquons deux profils distincts pour l'entraînement et la validation du modèle ARX. Ces profils, basés sur les mêmes principes expliqués précédemment, sont employés pour estimer la température mesurée par les capteurs CTN à partir des données de puissance appliquée aux puces.

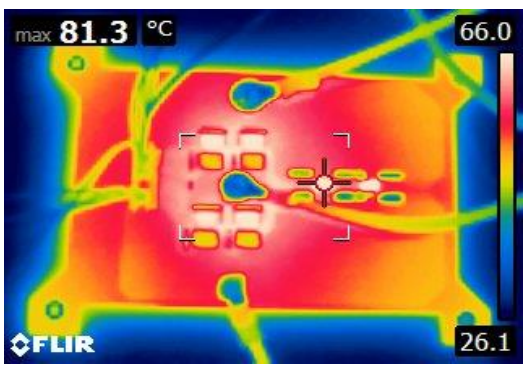

**Figure 3 : Carte 1 avec puces éloignées**

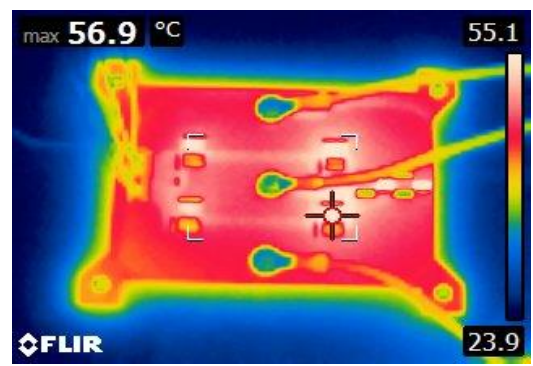

**Figure 4 : Carte 2 avec puces proches**

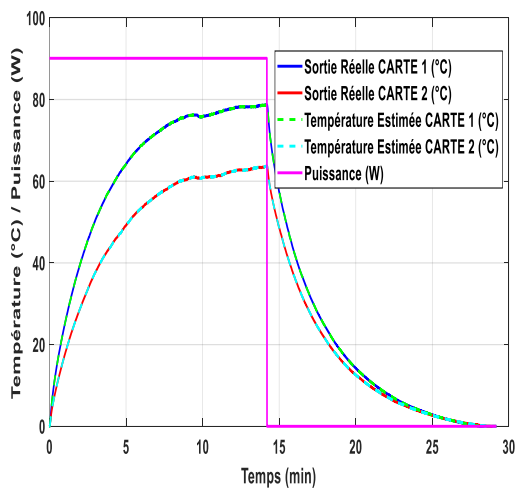

**Figure 5 : Identification du modèle ARX**

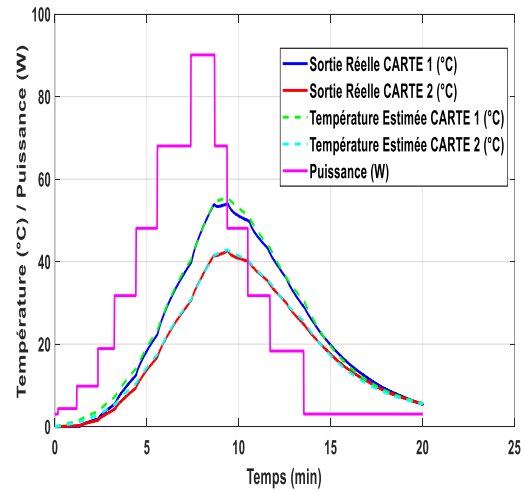

**Figure 6 : Validation du modèle ARX profil 1**

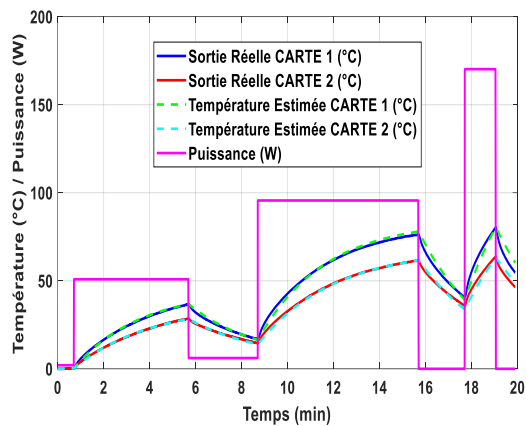
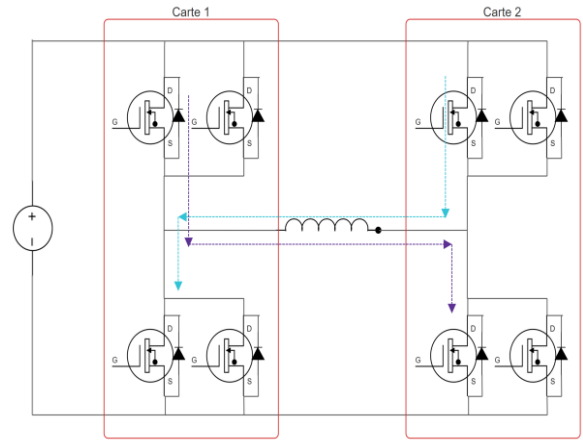

Figure 7 : Validation du modèle ARX profil 2          Figure 8 : Schéma électrique du pont en H

D'après l'analyse des figures 3 et 4, il est clair que la carte avec des puces plus rapprochées présente un point chaud significativement plus élevé par rapport à l'autre carte, avec une différence de température d'environ 20°C. Les figures 5,6 et 7, qui illustrent la variation de la température en fonction du temps pour les deux capteurs CTN selon les profils de puissance utilisés durant les phases d'entraînement et de validation, confirment également cette augmentation de température.

L'application de notre modèle ARX ajusté par SVD a donné des résultats remarquables, avec un ajustement de près de 100 % lors de la phase d'entraînement, validant sa capacité à suivre les variations de température avec une grande précision. Pour la phase de validation, un profil de puissance testant de multiples points de fonctionnement (figure 6) a maintenu un ajustement élevé à 96 %. Pour un second profil (figure 7), comportant moins de points de fonctionnement mais maintenus sur une durée plus longue, a légèrement réduit l'ajustement à 94 %. Ce dernier cas, représentant plus le régime permanent, démontre la fiabilité du modèle dans la durée. Ces constatations confirment l'efficacité du modèle ARX pour une estimation précise des conditions thermiques dans divers scénarios opérationnels.

## 4. Conclusion

Notre étude démontre l'efficacité de l'approche ARX, combinée avec la régularisation SVD, pour la prédiction thermique dans l'électronique de puissance. Elle souligne la précision des modèles ARX dans l'estimation de la température de jonction des semi-conducteurs WBG, en se basant uniquement sur des données expérimentales. Cette méthode se distingue comme une alternative robuste et fiable aux modèles thermiques RC, évitant la nécessité d'une compréhension exhaustive des propriétés matérielles. La contribution significative de cette recherche réside dans sa capacité à affiner les prédictions thermiques, essentielle pour le développement et l'optimisation des dispositifs de puissance.

## Références